\documentclass[%
 reprint,
amsmath,amssymb,
 aps,
]{revtex4-2}
\usepackage[english]{babel}
\usepackage{graphicx}
\usepackage{dcolumn}
\usepackage{bm}
\usepackage{multirow}
\usepackage{bm}
\usepackage{epsfig}
\usepackage{amsmath}
\usepackage{amssymb}
\usepackage{gensymb}
\usepackage[usenames]{color}
\usepackage{color}
\usepackage{epstopdf}
\usepackage{float}
\usepackage{caption}
\usepackage{subcaption}

\usepackage{xcolor}
\usepackage{ulem}
\usepackage{hyperref}

\begin{document}

\title{Understanding the Interlayer Coupling in\\ 1T/1H-NbSe$_2$ Hetero-Bilayers}

\author{Roman Pico}.
 \email{pico@ifir-conicet.gov.ar}
\author{Paula Abufager}
\email{abufager@ifir-conicet.gov.ar}
\author{Ignacio Hamad}

\affiliation{%
 Instituto de F\'{\i}sica de Rosario, Consejo Nacional de 
Investigaciones Cient\'{\i}ficas y T\'ecnicas (CONICET) and
Facultad de Ciencias Exactas, Ingenier\'{\i}a y Agrimensura.  
Universidad Nacional de Rosario, 27 de Febrero 210 Bis (2000) Rosario, Argentina 
}%

\author{Roberto Robles}
\author{Nicolas Lorente }
\affiliation{Centro de Física de Materiales CFM/MPC (CSIC-UPV/EHU), Paseo Manuel de Lardizabal 5, San Sebastián 20018, España}

\date{\today}

\begin{abstract}
The properties of 2D materials are strongly influenced by their substrate, leading to a variety of "proximity effects" like screening, charge transfer, and hybridization. Surprisingly, there is a dearth of theoretical studies on these effects. Particularly, previous theoretical research on the Star of David (SOD) structure in 1T-NbSe$_2$ has focused on single-layer configurations or stacking with the same 1T phase without any real substrate. Here, we depart from these approaches and explore how these proximity effects shape the electronic and magnetic properties of the 1T-NbSe$_2$ phase when it is grown on the metallic 1H-NbSe$_2$ substrate.
Using Density Functional Calculations, we establish a common framework to define the key characteristics of both free-standning 1T-NbSe$_2$ and 1H-NbSe$_2$. We then identify the optimal stacking arrangement for these two layers, revealing a transfer from the 1T to the 1H phase and a reorganization of charge within each layer. Our findings indicate that the magnetic moment of the SOD structure is still robust; however, is diminished due to a reduction in the on-site Coulomb interaction of the Hubbard bands. Additionally, the interlayer coupling induces metallicity in the 1T phase and increases the decoupling of the lower Hubbard band from the valence band.

\end{abstract}

\maketitle

\section{Introduction}

In recent years, two-dimensional (2D) van der Waals (vdW) materials, particularly transition metal dichalcogenides (TMDs), have captured the attention of the scientific community\cite{vdwmaterials-1,vdwmaterials-2}. These materials, heralded for their versatile applications in optoelectronic and spintronic technologies, offer a unique playground for material engineering and device fabrication\cite{2d-properties-1}. Among TMDs, NbSe$_2$ is a prominent example, showcasing a spectrum of electronic phases such as charge density wave (CDW), superconductivity, and variable electronic structures in its different phases like single layers (SL) of the 1H and 1T phases.\cite{ising-pairing.nbse2,Ugeda2016-naturephysics-sl-nbse2,2h-nbse2-cdw-ugeda-robles,cdw-sc-2d-nbse2-1h,Liu2021}. The 1T phase of NbSe$_2$, distinct from the 1H phase in terms of symmetry and electronic properties, adds an extra layer of complexity and potential to these materials, paving the way for tailoring the overall properties of the system through phase engineering.  
While the 1H phase is a metal, that at low temperatures  ($T_c$ = 7.2K for bulk and Tc = 3.2K for the monolayer \cite{nbse2-2h-tc-1972-bulk, NbSe2-2h-Tc-2009,ising-pairing.nbse2,Anisotropic1H-1,anisotropic1H-2}) develops superconductivity, the 1T was first identified as a Mott
insulator\cite{npg2016mott}, recently instead it has been characterized as a Charge Transfer Insulator (CT)\cite{Liu2021}.

The interplay between the 1H and 1T phases in hetero-layers of NbSe$_2$ enriches its physical properties, allowing for the manipulation of electronic and spin states in novel ways. This ability to tailor the characteristics of TMDs by controlling their phase composition opens new avenues for customized material properties, particularly in the context of proximity effects in heterostructures and their applications in advanced technologies \cite{Liu2023d}.

More recently, magnetic impurities have been deposited on clean surfaces of NbSe$_2$, resulting in the appearance of Yu-Shiba-Rusinov (YSR) bound states in the superconducting
gap\cite{franke-2019-nbse2-fe}. This could be linked to the creation of a topological phase and Majorana bound states. The experimental exploration of the vertically stacked 1T and 1H  NbSe$_2$ layers (hereafter referred to as 1T/1H) has unveiled intriguing phenomena that occur at the interface of these two phases. 
When the 1T phase is combined with the 1H phase, distinct and noteworthy effects have been observed, such as the emergence of Yu-Shiba-Rusinov (YSR) bound states and Kondo resonances within the superconducting gap\cite{Liu2021}. These effects highlight the complex and unique interactions at the 1T/1H interface, showcasing how the combination of different phases can lead to new physical behaviors. 

These experimental results suggest a significant modification of electronic and magnetic properties at the interface. This modification indicates an underlying influence that could be akin to a proximity effect\cite{1t-1t-interaction}, where the presence of one phase impacts the properties of the adjacent phase. These findings are pivotal in understanding the coupled dynamics of the 1T and 1H phases and offer valuable insights into the potential mechanisms driving their interaction. This experimental backdrop forms an important context for our theoretical study.

From the theoretical point of view and within the framework of calculations based on the Density Functional Theory (DFT), the simultaneous description of monolayers in 1H and 1T phases of $NbSe_2$ is a great challenge due to the different physical characteristics that each of them presents. 
Consequently, different approximations are utilized in their computational treatment to address this complexity and enhance the overall accuracy of the theoretical framework \cite{Divilov_2021,2h-nbse2-cdw-ugeda-robles,Calandra,Liu2021}. 

In this theoretical work, we intend to simultaneously capture the main features of both phases from DFT calculations finding a common set of parameters such as lattice parameter and effective on-site Coulomb interaction (U$_{eff}$), that correctly describe both phases. Then we shift our focus towards exploring the interactions at the interlayer interface. This exploration is crucial for understanding how the distinct properties of each phase influence each other when combined in a heterostructure. We investigate the nuances of interlayer coupling, particularly how it affects the electronic and magnetic properties of the combined system. This part of our work aims to provide a deeper insight into the synergistic effects at play between the 1H and 1T phases, shedding light on the potential of these interactions to induce novel phenomena in NbSe$_2$ heterostructures.

\section{Methods}

Spin polarized DFT calculations were performed with the VASP code \cite{Kresse1993a,Kresse1993b,Kresse1996a,Kresse1996b,Kresse1999,Hafner2008} within the slab-supercell approach and using the projector augmented-wave (PAW) method \cite{Kresse1999}.
Wave functions were expanded using a plane wave basis set with an energy cut-off of 500 eV. 
 We used the PBE functional \cite{Perdew1996} to treat the exchange-correlation energy. We
added the missing van der Waals interactions using the Tkatchenko-Scheffler scheme \cite{Tkatchenko2009}. The strongly-correlated correction is considered with the DFT+U approach \cite{Dudarev1998} to deal with the Nb d-electrons.  In this study, we investigated the influence of varying U$_{eff}$ values for Nb(d) atoms on the lattice constant, electronic properties, and magnetic properties of isolated 1T and 1H monolayers (see Supporting Information). Furthermore, we used linear response theory \cite{Cococcioni2005}  to estimate an optimal $U_{\text{eff}}$ value for both monolayers. Overall, this approach allowed us to analyze the interplay between electronic structure, magnetic behavior, and lattice parameters, providing valuable insights into the properties of these materials.
The isolated monolayers of 1H and 1T were studied using 1$\times$1 and $\sqrt{13}\times\sqrt{13}$ unit cells. Test calculations for the 3$\times$3 CDW of the 1H phase revealed that our primarily conclusions remain unchanged, even when considering a 1$\times$1 unit cell. The 3$\times$3 CDW in NbSe$_2$ is weak and does not significantly alter the band structure or DOS. For the heterobilayer 1T/1H, a $\sqrt{13} \times \sqrt{13}$ unit cell was used. It is worth mentioning that incorporating a commensurate unit cell for the CDW of both 1T and 1H layers would necessitate an unfeasible large unit cell. The 4p$^6$4d$^4$5s$^1$ states of Nb atoms and the 4s$^2$4p$^4$ states of Se atoms are treated as valence states.
For ionic relaxation, we used a  13$\times$13$\times$1 and 3$\times$3$\times$1  k-grids for 1$\times$1 and $\sqrt{13} \times \sqrt {13}$, respectively. For the PDOS, we used a  100$\times$100$\times$1 and 18$\times$18$\times$1  k-grids for 1$\times$1 and $\sqrt{13} \times \sqrt{13}$, respectively.

To analyze the charge transfer process, we employed the Bader scheme \cite{Yu20111} to calculate the charge in the 1T/1H, 1T and 1H optimal structures. In order to gain further insights, we also computed the difference between the charge densities of the entire 1T/1H model and the separated 1T and 1H layers, denoted as $\Delta \rho = \rho_{1T/1H} - \rho_{1T} -\rho_{1H}$. Here, $\rho_{1T/1H}$, $\rho_{1T}$, and $\rho_{1H}$ refer to the charge densities of the whole 1T/1H system, the 1T layer, and the 1H layer, respectively. To obtain $\rho_{1T}$ and $\rho_{1H}$, we removed the 1H monolayer and the 1T layers from the entire system respectively, while keeping the remaining atomic structures unchanged.

\section{Results and Discussion}

\subsection{Single-layer (SL) of 1T and 1H NbSe$_2$}

\begin{figure*}[ht!]          
\begin{center}
\includegraphics*[width=1.6\columnwidth]{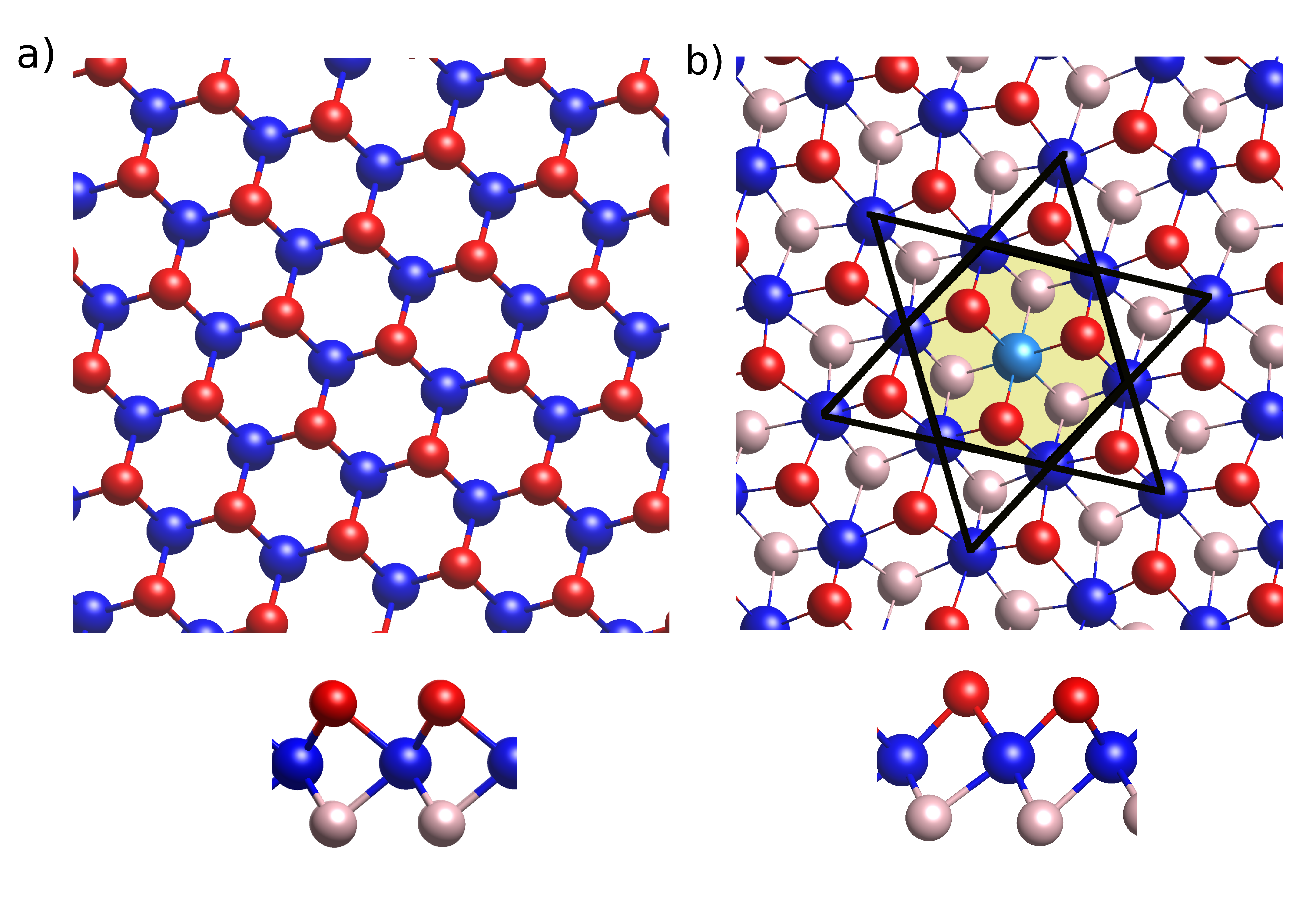} 
\end{center}
\caption{Structure of (a)  1H and (b) 1T single-layer configurations of $NbSe_2$. In these diagrams, blue spheres represent Nb atoms and  red (pink) spheres represent Se atoms in the top (bottom) layers of the arrangement. The central region (CR) of the SOD within the 1T structure is indicated by the yellow shading. This region prominently features the central Nb atom (Nb$_c$), depicted as a cyan sphere, which is at the heart of the SOD. Surrounding Nb atoms that are in close proximity to Nb$_c$ are represented as blue spheres (hereafter referred to as Nb$_1$). The pink Se atoms inside this area are named, $Se_c$ and the ones outside are $Se_o$ (see Figure S1 in the SI for visualization of the $\sqrt{13}\times \sqrt{13}$ unit cell of the 1T structure).}
\label{geom}
\end{figure*}

Single layers of  1H-NbSe$_2$  consist of hexagonal arrangements of Nb atoms enveloped within a trigonal prismatic Se atom environment (see Figure \ref{geom}). Prior theoretical investigations have highlighted various atomic structures that are compatible with the 3$\times$3 CDW distortions, and these alternatives exist within a narrow energy range of approximately 3 meV per formula unit \cite{2h-nbse2-cdw-ugeda-robles}. This intriguing observation suggests a coexistence of these structures. Moreover, the deviations in atomic positions from the ideal 1$\times$1 arrangement are relatively minor \cite{2h-nbse2-cdw-ugeda-robles}. These findings collectively illuminate the complex interplay between structural modifications and electronic behaviours within the intricate landscape of 1H-NbSe$_2$. When it comes to the role of magnetism in 1H-NbSe$_2$, the exact nature of this phase is not clear \cite{Divilov_2021, 2h-nbse2-cdw-ugeda-robles, Wickramaratne2020}.

Regarding the  1T-NbSe$_2$ phase, mono- and few-layers are, so far, experimentally accessible using molecular beam epitaxy  \cite{npg2016mott, Liu2021,twistronics,mott-colapse,Liu2023}. The single-layer (SL) 1T-NbSe$_2$ phase is characterized by a distinctive star-of-David (SOD) CDW pattern with a $\sqrt{13} \times \sqrt{13}$ periodicity \cite{npg2016mott}. Within this CDW pattern, each SOD comprises twelve outer Nb atoms that undergo contraction toward a central Nb atom. Additionally, six Se atoms are positioned in the upper plane and six in the lower plane of each SOD, as visually depicted in Figure \ref{geom}.  In contrast to the 1H polytype, in the 1T phase, each Nb atom achieves octahedral coordination with its neighboring Se atoms. 1T-NbSe$_2$ is proposed as a Charge Transfer insulator \cite{Liu2021}, in which the Valence Band (VB), mainly composed of the Se p$_z$ states, appreciably hybridizes with the Lower Hubbard Band (LHB) \cite{Liu2021b,twistronics}.

In light of our ultimate objective, which centers on the comprehensive investigation of the physical characteristics exhibited by bilayer configurations composed of 1T and 1H stacked structures, it becomes paramount to establish a singular computational approach capable of encapsulating their salient attributes. Of particular significance is the determination of a unique lattice constant and the appropriate choice of the $U_{\text{eff}}$ parameter for the treatment of Nb(4d) electrons. These parameters are pivotal not only in rendering an accurate representation of the metallic and insulating characteristics intrinsic to 1H and 1T single layers but also in elucidating their magnetic properties.

\begin{figure*}[ht!]                                        \begin{center}
\includegraphics*[width=1.6\columnwidth]{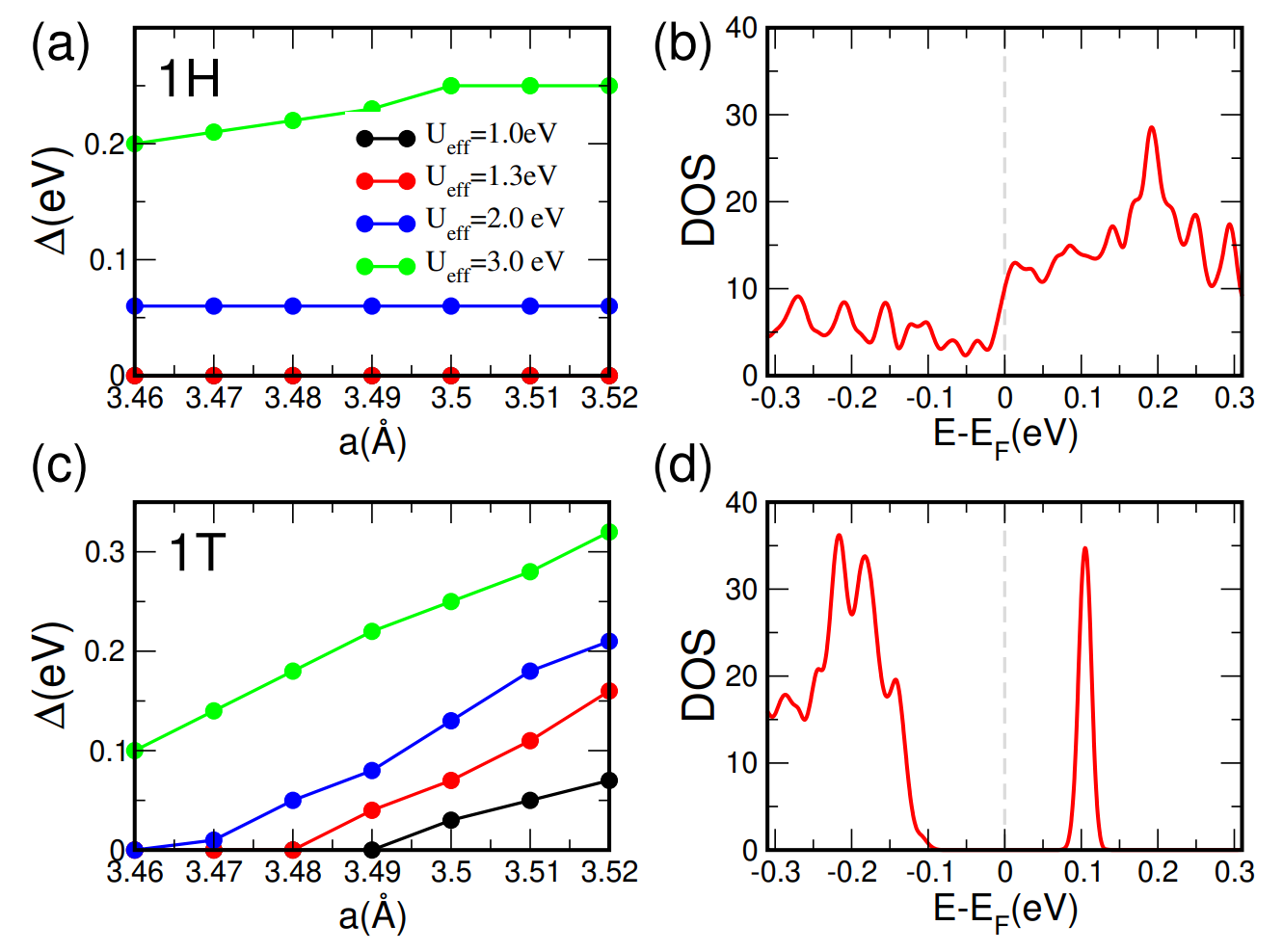} 
\end{center}
\caption{Effect of the lattice parameter, a (\AA), on the Band gap $\Delta$ (eV), of (a) 1H- and (c) 1T NbSe$_2$ monolayers with PBE+U (U$_{eff}$=1, 1.3, 2 and 3 eV). Density of states (DOS) calculated with  U$_{eff}$=1.3 eV and a=3.52 \AA~for (b) 1H- and (d) 1T- NbSe$_2$ monolayers. In (a) the red and black lines are overlapping. } 
\label{mag-y-dos}
\end{figure*}

For the 1H single layer (SL), the variation of the band gap, denoted as $\Delta$, with respect to the lattice constant is depicted in Figure \ref{mag-y-dos}(a) for different values of $U_{\text{eff}}$.  As expected, for high enough values of the Hubbard parameter, $U_{\text{eff}}  \geq$ 2 eV, the system becomes an insulator. Additionally, due to the presence of an odd number of electrons within the unit cell, the total magnetic moment is M$_T$ = 1$\mu_B$ when the system becomes an insulator. An intriguing observation to note is that, for a fixed value of $U_{\text{eff}}$, the strain applied to the SL does not exert a pronounced influence on the band gap. All in all, the metal-to-insulator transition triggered by strong electron-electron correlations imposes constraints on the permissible range of $U_{\text{eff}}$ values that we can select to accurately model the established electronic characteristics of the 1H SL. In particular, when considering $U_{\text{eff}}=$1.3 eV, it is noteworthy that the SL exhibits a consistently metallic behaviour across the entire range of lattice parameters examined. 
While the presented results correspond to a 1$\times$1 structural model, test calculations considering geometries based on the 3$\times$3 CDW  exhibit similar trends, further enhancing the validation of our findings. 

In the context of the 1T SL, we apply a methodology similar to that used for the 1H counterpart. However, we investigate a $\sqrt{13} \times \sqrt{13}$ unit cell configuration, which is a distinctive feature associated with the SOD. 

In Figure \ref{mag-y-dos}(c), we illustrate a pronounced dependence of $\Delta$ on both lattice parameters and the $U_{eff}$  interaction parameter. This dependence also translates to the magnetic properties  (see Figure S2 in the SI). For instance, for small enough values of both parameters, the known insulator state and the magnetic characteristics of the 1T phase (SOD) are lost. 
This outcome aligns with previous research findings that have underscored a strong correlation between the charge transfer gap and the structural distortion linked to the SOD \cite{Liu2019}. Particularly, the authors have emphasized the potential of applying mechanical strain as a promising strategy for the precise adjustment and fine-tuning of the charge transfer gap in this material \cite{Liu2019}. It is noteworthy to highlight that, within the context of the 1T SL, an increase in strain results in a corresponding augmentation of the magnetic moment and this phenomenon may be attributed to the enhanced super-exchange interaction, as suggested by Liu et al. \cite{Liu2023b}. 

Putting together the information gathered so far on 1H and 1T SLs, it is clear that our nuanced and balanced approach can effectively capture and describe the collective physical properties of both SL configurations. In particular, we have identified that setting the effective Hubbard interaction parameter, $U_{eff}$, to 1.3 eV, and the lattice constant to 3.52 \AA, successfully reproduces the main characteristics of the density of states (DOS) (see \ref{mag-y-dos}(b-d)) and magnetic characteristics of both the experimental 1T and 1H phases. This choice preserves the metallic and insulating properties characteristic of each phase, respectively. For the 1T-SL, our calculations indicate a band gap of $\approx$0.15 eV, in good agreement with the experimental gap (i.e 0.15 eV) found by Liu et al \cite{Liu2021}. Furthermore, the SOD  exhibits behavior akin to that of a spin-1/2 system, with the central Nb atom ($Nb_c$) and its nearest neighbor Nb atom ($Nb_1$) displaying magnetic moments of $m_{Nb_c}=$0.43 $\mu_B$ and $m_{Nb_1}=$0.36 $\mu_B$ (in total), respectively.

It is worth noting that the selected lattice constant exhibits minimal dispersion, with a variation of less than 1\% compared to the optimal values, being $3.49 \textup{\r{A}}$ for 1H and $3.50 \textup{\r{A}}$ for 1T. Furthermore, these chosen lattice constants align well with experimental measurements (i.e. 348$\pm$14 pm and 343$\pm$11 pm for 1H and 1T respectively) \cite{Huang2022}.

Additionally, we implemented linear response theory \cite{Cococcioni2005} and obtained for both phases an $U_{eff}=1.1$ eV, very close to our choice of $U_{eff}$. This validates both the chosen value and the use of a single one for both phases.

\subsection{Heterogeneous 1T/1H NbSe$_2$ bilayer}

\begin{figure*}[ht!]                                        \begin{center}
\includegraphics*[width=1.8\columnwidth]{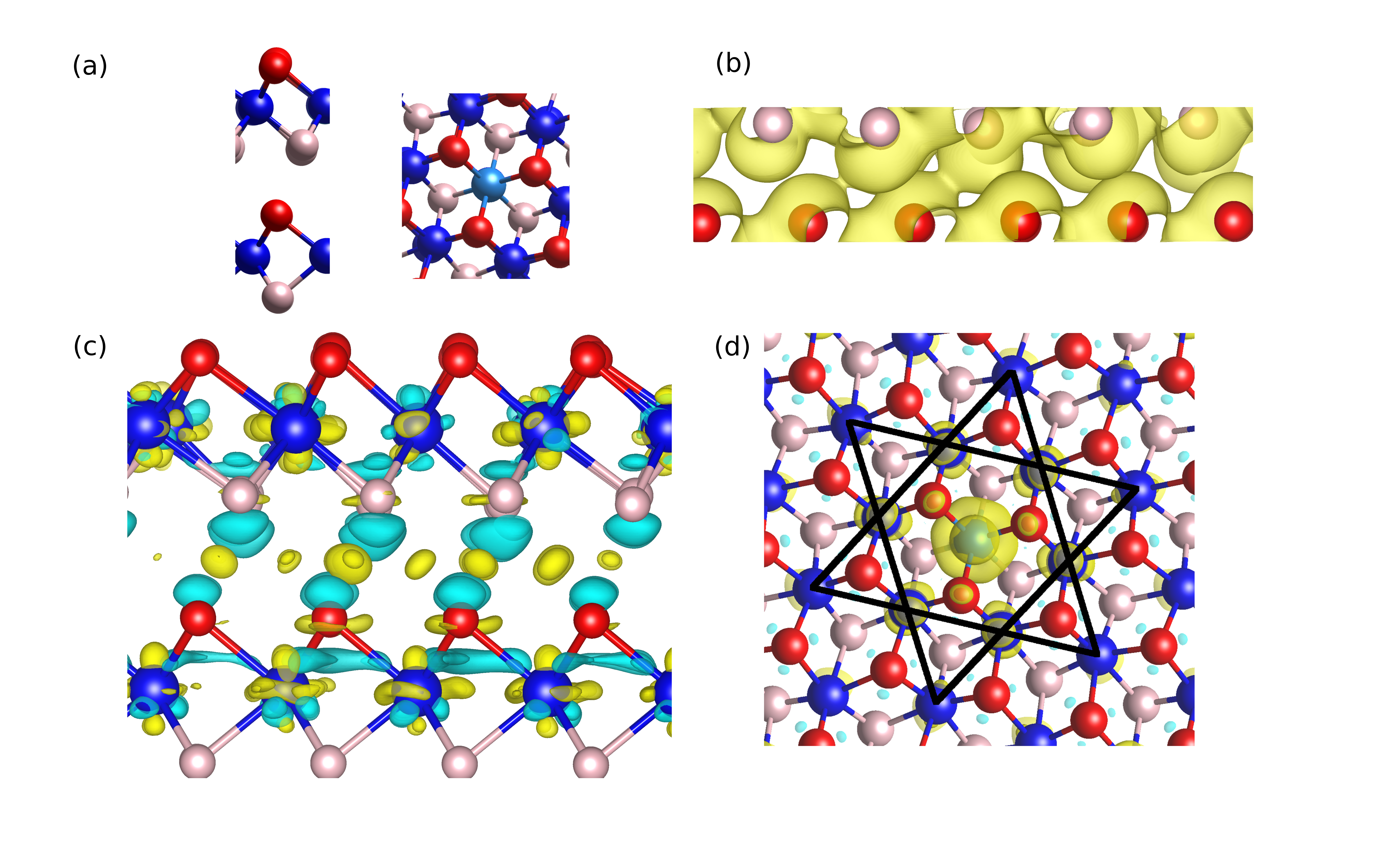} 
\end{center}
\caption{Optimal Stacking for the 1T-1H Bilayer:
(a) schematic representation of the geometrical arrangement. In this diagram, Se atoms in the 1TB layer are depicted as pink balls, residing in the hollow sites of the Se atoms in the 1HT layer, shown as red balls. Additionally, Nb atoms in the 1T layer are positioned directly above the Nb atoms in the 1H layer. (b) Charge Density at the Interface between the Se 1T and 1H Layers in the Valence Band (VB), with an isovalue of  10$^{-5}$ $e^{-}/\AA^3$). (c) Induced Charge Density, $\Delta \rho = \rho_{1T-1H}-\rho_{1T}-\rho_{1H}$, representing the difference in charge density between the 1T-1H bilayer and the individual 1T and 1H layers (isovalue= 7 $\times$ 10$^{-4}$ $e^{-}/\AA^3$). (d) Spin density distribution within the 1T section of the bilayer (isovalue= 1.5 $\times$ 10$^{-3}$ $e^{-}/\AA^3$).} 
\label{bilayer}
\end{figure*}

To determine the optimal arrangement of the 1T-NbSe$_2$ on 1H-NbSe$_2$ substrate, we explored different stacking configurations (Figure S3 in the SI). There are six high symmetry adsorption sites where the 1T monolayer can stack on top of the 1H monolayer \cite{ugeda-vse2-stacking,stackings-MoSe2}. Figure \ref{bilayer}(a) shows the lowest energy stacking geometry. Here, the Nb atoms in the 1T and 1H SLs lie on top of each other, and the interfacial Se atoms in the bottom layer, denoted as B1T within the 1T arrangement, align precisely with the hollow sites corresponding to the interfacial Se atoms in the top layer, designated as T1H within the 1H structure. This geometric arrangement gives rise to a hexagonal pattern when viewed from the top. The interlayer distance between the Nb atoms in 1T and 1H layers is d$_{int}$= 6.08 \AA\space consistent with the distance between planes in bulk 2H-$NbSe_2$  (i.e.  6.2 $\pm$ 0.1 \AA) \cite{Zhang2022} while the distance between Se in the B1T and T1H layers is 2.8 \AA. The cohesive energy between the layers is -4.47 eV, with the van der Waals (vdW) contribution being -5.13 eV.

 The hexagonal stacking among Se atoms in the interface greatly enhances the interfacial wavefunction overlap among them. Each Se atom within the B1T layer engages in hybridized interactions with its closest neighbors in the T1H layer, as visually depicted in Figure \ref{bilayer}(b). This bonding interaction results in the emergence of interlayer bonds that exhibit characteristics akin to "covalent-like quasi-bonds."\cite{Dai2023} Such interlayer bonding behavior has been observed in other bilayer systems as well, as referenced in prior works \cite{Dai2023}. These results serve as a distinctive hallmark of the interlayer interactions occurring between the 1T and 1H layers. 

The interlayer interaction is thoroughly elucidated through a comprehensive analysis of the charge transfer process between the 1T and 1H arrangements, as visualized in Figure \ref{bilayer}(c) and summarized in Table \ref{transf-carga}. This investigation reveals a charge transfer and reorganization when comparing the monolayers to the newly constructed bilayer structure. In particular, the central Nb$_c$ atom within the 1T phase, and to a somewhat lesser extent, the neighboring Nb$_1$ atoms, exhibit a marginal decrease in charge, with charge variations per atom quantifying at less than 0.01 $e$. The external Nb atoms of the SOD accumulate a marginal charge of $\approx$ 0.01 $e$ per atom. All in all, the net change in the overall charge of the Nb layer within the 1T phase amounts to a mere increment of $\approx$ 0.01 $e$. Notably, the lower Se atoms within the 1T phase experience a charge loss approximately five times more significant than their upper counterparts within the same phase, resulting in a net charge reduction of 0.18 $e$ within the Se layers of the 1T phase. Consequently, a charge transfer of 0.17 $e$ is observed, flowing from the 1T phase to the 1H phase. This is in line with our expectations, as it is consistent with the reduced work function of the 1T phase, $W_f^{1T} = 5.21$ eV, relative to the 1H phase $W_f^{1H} = 5.36$ eV. Within the 1H layer, a distinct redistribution of charge is also observed. The Nb layer increases its charge by 0.27 $e$, while there is a slight charge depletion (0.10 $e$) in the Se layers. It is important to highlight that the charge transfer from the 1T phase to the 1H layer is significantly reduced when compared to the TaS$_2$ system, as previously documented in Crippa et al. \cite{crippa2024}, and this variation in the charge transfer process may lead to other distinct behaviors and outcomes.

All in all, the charge transfer and reorganization processes, as outlined earlier, result in the concentration of charge at the interface between the 1T and 1H layers, with a notable emphasis on the vicinity of the Se atoms of the B1T and T1H layers (see Figure\ref{bilayer}(c)). This observation aligns with the presence of interfacial states, visually exemplified in Figure \ref{bilayer}(b). Additionally, when the 1T and 1H layers come into close proximity, it leads to a reduction in the magnetic moment of the SOD structure, as indicated in Table \ref{transf-carga}. Similar to the single-layer (SL) case, the magnetic moment of the SOD (see Figure \ref{bilayer}(d)), primarily originates from the Nb(dz$^2$) states of Nb$_c$ and Nb$_1$ atoms. However, it is noteworthy that the total magnetic moment of these states decreases from 0.72 $\mu_B$ in the 1T-SL configuration to 0.43 $\mu_B$ in the 1T-BL configuration. 

\begin{table*}
	\begin{tabular}{ p{3cm} p{3.0cm}   p{3.0cm} p{3.0cm} p{3.0cm}  } \\ 		\hline
		\hline
		Species & $\Delta_q$(Nb) (e$^{-}$) & $\Delta_q$(Se) (e$^{-}$) & $\Delta_m$(Nb) ($\mu_B$) &  $\Delta_m$(Se) ($\mu_B$)  \\
		\hline
         1T  & 0.014 (tot.) & -0.180 (tot.) &  -0.180 (Nb$_c$) & 0 (av.)            \\
         \hline 
         \hline
         1H  & 0.273 (total) & -0.107 (total) & -0.102 (av.)& 0 (av.)   \\
        \hline
		\hline
	\end{tabular}
\let\nobreakspace\relax
\caption{Bader charge transfer analysis was performed for the 1T/1H bilayer, yielding values of $\Delta_q$(Nb) and $\Delta_q$(Se), which represent the total charge transferred for all Nb and Se atoms, respectively, in either the 1T or 1H component of the hetero-bilayer relative to the isolated monolayers. Additionally, the magnetic moment changes in Nb and Se atoms for the 1T or 1H components of the bilayer with respect to the isolated monolayers are represented by $\Delta_m$(Nb) ($\mu_B$) and $\Delta_m$ ($\mu_B$), respectively. The value corresponding to the center of the SOD is given for the 1T component, while the average value for all atoms is given for the other cases.}
\label{transf-carga}
\end{table*}

The electronic hybridization of interfacial Se atoms and the charge transfer processes among the layers are pivotal factors that significantly influence the electronic structures of the entire system, contrasting with the properties of their isolated components.

\begin{figure*}[ht!]                                        \begin{center}
\includegraphics*[width=1.8\columnwidth]{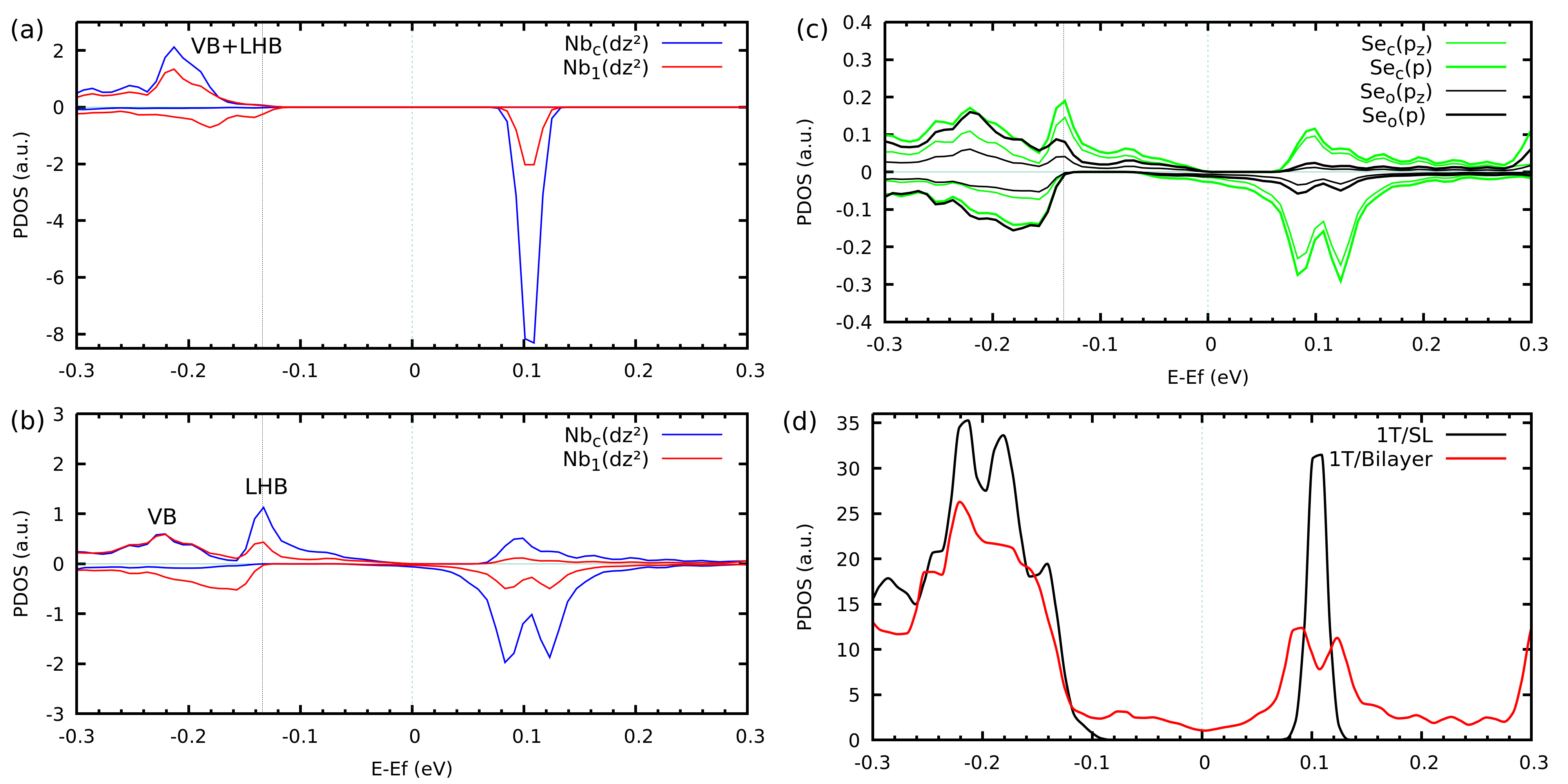} 
\end{center}
\caption{Projected Density of States (PDOS) into the Nb$_c$(dz$^2$) (in blue) and Nb$_1$(dz$^2$) (in red) states in both the single layer (a) and bilayer (b) configurations. Additionally, (c) illustrates the PDOS into the Se$_c$(p) (thick green line), Se$_c$(pz) (thin green line), Se$_o$(p) (thick black line), and Se$_o$(pz) (thin black line) states. In the figures (a-c), the reported data is normalized per atom. (d) Full PDOS encompassing all atoms within the 1T-SL (in black) and the 1T part of the bilayer (in red). } 
\label{dos-y-induced-charge.jpg}
\end{figure*}

In the SL-1T, as depicted in Figure \ref{dos-y-induced-charge.jpg}(a), the valence band and the Lower Hubbard Band (LHB) are known to undergo hybridization, resulting in a broad peak \cite{liu-1t-SOD}, which, in our specific study, is centered at -0.22 eV. This hybridization arises from the significant overlap of the Se p$_z$ states within the valence band with the LHB \cite{liu-1t-SOD}. The complex interplay of these electronic states poses a substantial challenge when it comes to precisely determining the position of the LHB.  In contrast, the Upper Hubbard Band (UHB) in the 1T-SL exhibits a well-defined, narrow peak, which, in our investigation, is centered at 0.11 eV.

When we contrast the SL-1T with its bilayer counterpart (as depicted in Figure \ref{dos-y-induced-charge.jpg}b), the previously observed broad peak al -0.22~eV undergoes a notable transformation, resolving into two distinct smaller peaks. The peak at -0.14 eV  displays clear spin-polarized characteristics, primarily stemming from the involvement of Nb($d_{z^2}$) and Se($p_z$) orbitals (see Figure \ref{dos-y-induced-charge.jpg} b-c).
Notably, the contribution from the Nb$_c$ SOD center exceeds that of its nearest neighbor Nb atoms, Nb$_1$. These distinct features are also evident in the spatial distribution of the energy-resolved charge density (see Figure S4 in the SI). On the contrary, the peak positioned at -0.22 eV displays a diminished degree of spin polarization, embodying a more hybridized character. In this instance, Nb$_c$ and its nearest neighbor Nb atoms contribute almost equally, a trend consistently evident in the corresponding energy-resolved charge density  (refer to Figure S4 in the Supplementary Information). Notably, this particular electronic structure spans a broader energy range, with a distinct and substantial contribution originating from Se($p_z$) orbitals. All in all, our calculations suggest that the combined VB and LHB in the 1T-SL  undergo a distinct separation when the 1T layer is in close proximity to the 1H layer. Specifically, the peak centered at -0.14 eV predominantly exhibits LHB characteristics, while the structure with a central energy of -0.22 eV displays a more pronounced Valence Band (VB) character.

The UHB in our observations also undergoes substantial alterations when compared to the Single-Layer (SL) case. Notably, we have observed that the UHB peak broadens and separates into two distinct components. The exact cause of this splitting remains unclear, as both peaks exhibit significant contributions from Nb$_c$($d_{z^2}$) states, along with contributions from their nearest neighbors (as evidenced in Figure \ref{dos-y-induced-charge.jpg}b and further illustrated in Figure S4 in the Supplementary Information).

To further support the assignment of the LHB, VB and UHB we conducted a series of calculations using different U$_{eff}$ parameters within our DFT+U framework. These calculations provided valuable insights into the behavior of these bands. Specifically, we observed that the peak associated with the LHB shifts towards lower energies, and the gap between the LHB and UHB increases as the U$_{eff}$ values are elevated, in line with our expectations. Intriguingly, we also found that the splitting of the UHB is influenced by the U$_{eff}$ parameter (for detailed visualization, refer to Figure S5 in the Supplementary Information).

The preceding discussion highlights the significant impact of the proximity between the 1T and 1H layers on the behavior of the LHB and UHB, resulting in their shifting closer to the Fermi level.  This observed phenomenon can be attributed to the dielectric screening effects originating from the lower layer acting on the uppermost layer. This dielectric screening effectively reduces the on-site Coulomb repulsion (Hubbard U), thereby decreasing the energy gap between the LHB and the UHB\footnote{A similar reduction, attributed to the addition of extra layers, was also observed in  1T-TaSe$_2$\cite{1t-tase2-few-layers}.}. As a direct consequence of the reduction in U, a decrease in the charge within the majority spin is observed, along with an increase in the occupation of the Nb$_c$(dz$^2$) and Nb$_1$(dz$^2$) states in the minority spin. Importantly, this shift in charge distribution occurs while the overall total charge of the states remains relatively unchanged. This phenomenon serves to elucidate the underlying cause of the decrease in the magnetic moment of the SOD structure (additional details available in Figure S6 of the Supplementary Information).

The reduced on-site Coulomb repulsion U effectively decreases the gap between the Hubbard bands. In conjunction with their hybridization with the valence band, this leads to broadening, reduced localization, and extended tails. These effects contribute to the emergence of states at the Fermi level, consistent with the 1T phase's transition towards a metallic phase. Evidence of this is seen in the PDOS for the 1T whole arrangement, as depicted in Figure \ref{dos-y-induced-charge.jpg}d. The orbital composition of these states not only involves atoms and orbitals that are directly associated with the Hubbard bands. This observation strongly reinforces the idea that the overall system, rather than the Hubbard bands in isolation, plays a pivotal role in accounting for the population of these states and, consequently, for the insulating-to-metallic transition within the 1T phase in the bilayer system. This transition is a direct result of the proximity effect induced by the metallic 1H counterpart, reflecting the intricate interplay of electronic interactions within the system.

In summary, our findings indicate that the magnetic moment of the SOD persists within the heterostructure. Moreover, we find evidence of  hybridization between 1T and 1H layers, along with a  reduction in the effective Coulomb interaction, U, within the LHB and UHB.  These characteristics align well with the scenario proposed by Liu et al. \cite{Liu2021}, where itinerant electrons in metallic 1H-NbSe$_2$ couple with the spin of the SOD in 1T-NbSe$_2$, resulting in the emergence of the Kondo screening effect.

\section{Conclusions}

In this theoretical investigation, we conducted a comprehensive study of 1T-NbSe$_2$ and 1H-NbSe$_2$ SLs and their heterobilayer structure by means of Density Functional Theory. Our findings indicate that a $U_{eff}=$1.3 eV and a lattice constant of 3.52 \AA \space describe the overall physical properties of both SL configurations. For the 1T/1H heterobilayer, we explored the interlayer interaction between the SLs. Our investigation found a charge transfer and reorganization process between 1T and 1H layers of 0.17 $e$ flowing from the 1T phase to the 1H phase and supported by work function calculations. In particular, there is a concentration of charge at the interface between the 1T and 1H layers, primarily in the vicinity of the Se atoms. 
Upon bringing the 1T and 1H layers into close proximity, a discernible reduction in the magnetic moment of the star-of-David (SOD) structure is observed, diminishing from 0.72 $\mu_b$ to 0.43 $\mu_b$.
In the context of the 1T-SL, it is well-documented that the Valence Band (VB) and the Lower Hubbard Band (LHB) undergo hybridization, resulting in the formation of a broad peak. However, when the two SLs come into contact, this broad peak undergoes a notable transformation, splitting into two distinct structures. One is situated at -0.14 eV, characterized by clear LHB attributes, while the other, positioned at -0.22 eV, exhibits a diminished degree of spin polarization and a more hybridized character. The Upper Hubbard Band (UHB), which maintains a narrower profile in the SL, undergoes peak broadening while preserving its spin-polarized characteristics. This transformation of the electronic structure near the Fermi level is accompanied by a reduction in the Hubbard U, thanks to the dielectric screening effect between the layers. This reduction leads to an Insulator-to-Metal transition.
In summary, the reduction in the effective Hubbard U and the broadening resulting from hybridization between single layers (SLs) with extended tails  play a major role and provide a comprehensive explanation for the appearance of states at the Fermi level.
\begin{acknowledgments}
R.P., I. H. and P.A. acknowledge the financial support from CONICET PIP-0083 and the computational time provided by the CCT-Rosario computational center and Computational Simulation Center (CSC) for Technological Applications, members of the High Performance Computing National System (SNCAD-MincyT Argentina). 
This work was also supported by the project  PID2021-127917NB-I00 funded by MCIN/AEI/10.13039/501100011033, IT-1527-22 funded by the Basque Government, and 202260I187 funded by CSIC. 
\end{acknowledgments}

\section*{Data availability statement}
All data that support the findings of this study are included within the article (and any supplementary files).
%

\end{document}